\documentclass[11pt]{article}

\usepackage[T1]{fontenc}
\usepackage[utf8]{inputenc}
\usepackage{lmodern}
\usepackage{amsmath}
\usepackage{amssymb}
\usepackage{booktabs}
\usepackage{array}
\usepackage{tabularx}
\usepackage[margin=1in]{geometry}
\usepackage{xcolor}
\usepackage{listings}
\usepackage{microtype}
\usepackage[hidelinks]{hyperref}

\newcolumntype{Y}{>{\raggedright\arraybackslash}X}
\newcommand{\MatMul}{\mathcal{M}_{\langle 3,3,3\rangle}}
\newcommand{\ind}{\mathbf{1}}
\newcommand{\artifactrepo}{%
  \url{https://github.com/trylogical/cn122_add55}%
}

\lstdefinestyle{certificate}{
  basicstyle=\ttfamily\footnotesize,
  columns=fullflexible,
  keepspaces=true,
  breaklines=true,
  frame=single,
  rulecolor=\color{black!20},
  xleftmargin=0.4em,
  xrightmargin=0.4em
}

\title{55 Additions Suffice for \(3\times3\) Matrix Multiplication at Rank 23}
\author{\textbf{Samurdhi Karunaratne} \qquad \textbf{Anushka Idamekorala}\\
  \href{https://research.logical.io}{Logical AI}\\
  Correspondence: \href{mailto:sam@logical.io}{\texttt{sam@logical.io}}}
  
\date{July 24, 2026}

\begin{document}

\maketitle

\begin{abstract}
We give a 55-addition realization of rank-23 multiplication of two arbitrary
\(3\times3\) matrices. Together with its 23 bilinear products, the circuit
uses 78 scalar operations. This improves the previous state of the art of
56 additions, due to Sun. The construction starts from Perminov's public
58-addition realization \texttt{cr58\_cn122} of a ternary tensor; the
contribution is a shorter and, for this fixed orientation of that tensor,
provably optimal linear circuit: 13 additions on the left input, 14 on the
right input, and 28 at the output. The last circuit is obtained by
transposing a 14-addition factor circuit. Because the coefficient alphabet
is \(\{-1,0,1\}\) and the order of every bilinear product is retained, the
algorithm applies over every associative ring, commutative or not. We print
the full straight-line program and tensor factors and provide four exact
computational checks, including independent Python and Node.js
implementations of all 729 Brent identities over \(\mathbb Z\).

\medskip
\noindent\textbf{Keywords.}
Fast matrix multiplication; bilinear algorithms; tensor rank;
additive complexity; transposition principle; exact verification.
\end{abstract}

\section{Introduction}

Strassen's algorithm showed that matrix multiplication can be accelerated by
reducing the number of bilinear products rather than evaluating the defining
sums directly~\cite{strassen}. For \(3\times3\) matrix multiplication over
noncommutative rings, Laderman found a rank-23 algorithm in
1976~\cite{laderman}. Rank 23 remains the smallest known rank for general
\(3\times3\) multiplication, while recent work has substantially reduced the
number of additions required to realize rank-23
schemes~\cite{smirnov,karstadt,beniamini,martensson-beast,stapleton,martensson-rank23,perminov,sun}.

The sequence relevant here is 60 additions in Stapleton's construction, 59
in Mårtensson--Wagner--Stapleton, 58 in Perminov, and 56 in
Sun~\cite{stapleton,martensson-rank23,perminov,sun}. The present construction
lowers the upper bound by one more addition. It does not alter the
multiplicative rank.

The underlying tensor is Perminov's publicly released
\texttt{cr58\_cn122} factorization~\cite{perminov,perminov-repo}. The tensor
itself is pre-existing; the contribution here is the 55-addition schedule
and its exact fixed-tensor optimality certificate in the binary
\(x+y\)/\(x-y\) straight-line model.

\medskip
\noindent\textbf{Theorem 1.}
\textit{Two arbitrary \(3\times3\) matrices over any associative ring can
be multiplied with 23 scalar multiplications and 55 additions/subtractions.}
\medskip

The count has a simple anatomy:
\begin{table}[htbp]
\centering
\begin{tabular}{@{}lcr@{}}
\toprule
Circuit component & Map & Additions\\
\midrule
Left preprocessing & \(9\to23\) & 13\\
Right preprocessing & \(9\to23\) & 14\\
Output reconstruction & \(23\to9\) & 28\\
\midrule
\textbf{Total linear work} & & \textbf{55}\\
Bilinear layer & 23 products & 23 multiplications\\
\bottomrule
\end{tabular}
\caption{Operation-count anatomy of the circuit.}
\label{tab:anatomy}
\end{table}

Section~\ref{sec:model} fixes the algebraic and cost conventions.
Section~\ref{sec:reduction} explains why the three linear costs are
\(13,14,28\), including the matching lower bound for this tensor.
Section~\ref{sec:algorithm} is the directly executable algorithm.
Section~\ref{sec:certificate} prints the tensor certificate in two forms,
and Section~\ref{sec:audit} records the exact audit trail.
Section~\ref{sec:ai} discloses the AI-assisted route by which the circuit
was found.

We do not claim that 55 additions is globally optimal among all rank-23
decompositions, nor that it is optimal in a different circuit model allowing
other primitive operations. To the best of our knowledge, 55 is the smallest
addition count presently known for a rank-23 algorithm in this model.

\section{Algebraic and circuit conventions}
\label{sec:model}

We use row-major coordinates:
\[
C=AB=
\begin{pmatrix}
C_0&C_1&C_2\\
C_3&C_4&C_5\\
C_6&C_7&C_8
\end{pmatrix}
=
\begin{pmatrix}
A_0&A_1&A_2\\
A_3&A_4&A_5\\
A_6&A_7&A_8
\end{pmatrix}
\begin{pmatrix}
B_0&B_1&B_2\\
B_3&B_4&B_5\\
B_6&B_7&B_8
\end{pmatrix}.
\]
Thus \(C_0=A_0B_0+A_1B_3+A_2B_6\), and similarly for the other entries.
Multiplication order is always a left input form in \(A\) times a right
input form in \(B\), so the construction is meaningful over noncommutative
coefficient rings.

A rank-\(R\) bilinear algorithm is a decomposition
\[
\MatMul=\sum_{r=0}^{R-1}U_r\otimes V_r\otimes W_r,
\]
where \(U_r,V_r,W_r\) are length-nine coefficient vectors. It computes
\[
M_r=
\left(\sum_{i=0}^{8}U_{r,i}A_i\right)
\left(\sum_{j=0}^{8}V_{r,j}B_j\right),
\qquad
C_k=\sum_{r=0}^{R-1}W_{r,k}M_r.
\]
The displayed scheme has \(R=23\), and all entries of \(U,V,W\) belong to
\(\{-1,0,1\}\).

The addition count is measured in the standard three-stage bilinear
straight-line-program model. A first linear SLP derives the 23 left forms
from entries of \(A\) only, and an independent linear SLP derives the 23
right forms from entries of \(B\) only. Corresponding forms are then paired
in the 23 products \(M_r\). A third linear SLP, whose inputs are those
products, reconstructs the nine entries of \(C\). Gates do not mix
\(A\)-derived and \(B\)-derived wires, and the reported additive cost is the
sum of the two preprocessing costs and the postprocessing cost. Within each
linear SLP, inputs are free, a gate computing \(x+y\) or \(x-y\) from
previously available quantities costs one, and copies and sign changes cost
zero. The 23 bilinear products are counted separately.

Correctness is equivalent to the Brent identities~\cite{brent}
\begin{equation}
\sum_{r=0}^{22}
U_{r,(i,k)}V_{r,(k',j)}W_{r,(i',j')}
=
\ind[i=i']\ind[j=j']\ind[k=k']
\label{eq:brent}
\end{equation}
for all \(i,j,k,i',j',k'\in\{0,1,2\}\). There are \(3^6=729\)
identities.

\section{From a 58-addition realization to a 55-addition circuit}
\label{sec:reduction}

Perminov's \texttt{cr58\_cn122}
factorization~\cite{perminov,perminov-repo}\footnote{The exact source is
\href{https://github.com/dronperminov/FastMatrixMultiplication/blob/98ba522db92b74f1f8c561a78038ff3091356d73/schemes/results/addition_reduced_ZT/3x3x3_m23_cr58_cn122_ZT_reduced.json}{%
\texttt{3x3x3\_m23\_cr58\_cn122\_ZT\_reduced.json}} at repository commit
\texttt{98ba522db92b74f1f8c561a78038ff3091356d73}; the companion repository
contains a byte-identical copy.}
realizes its three linear sides with costs \(14\), \(15\), and \(29\), for a
total of 58 additions. The third side is a 23-to-9 output map. Its transpose
is therefore a 9-to-23 factor map with a 15-addition realization.

\subsection{Exact synthesis of the factor maps}

We optimized each 9-to-23 factor map exactly in the binary
addition/subtraction model. Identify nonzero forms that differ only by sign,
and let \(d(F)\) be the number of distinct non-basis output directions of a
factor map \(F\). Every such direction must be created by at least one gate,
so \(C(F)\ge d(F)\). If equality held, every gate would have to create one
of the target directions: no auxiliary direction could occur.

For each factor, an exhaustive dependency search enumerated every way a
target direction can be obtained as a signed sum or difference of two basis
or previously available target directions. No topological schedule exists
using only \(d(F)\) gates. Hence \(C(F)\ge d(F)+1\). The explicit schedules
found at that bound establish Table~\ref{tab:factor-costs}.

\begin{table}[htbp]
\centering
\small
\begin{tabular}{@{}lrrrr@{}}
\toprule
Factor map & Source cost & \(d(F)\) &
\(d(F)\)-gate circuit? & Exact cost\\
\midrule
\(U:9\to23\) & 14 & 12 & No & 13\\
\(V:9\to23\) & 15 & 13 & No & 14\\
\(W:9\to23\) & 15 & 13 & No & 14\\
\bottomrule
\end{tabular}
\caption{Exact costs of the three 9-to-23 factor maps. ``No'' in the
fourth column means that exhaustive search found no circuit attaining the
direction-count floor, and hence \(C(F)\ge d(F)+1\).}
\label{tab:factor-costs}
\end{table}

\subsection{Transposing the output computation}

The 14-gate factor circuit for \(W\) is then transposed by reverse
accumulation. The linear-circuit transposition
principle~\cite{burgisser} converts an \(n\)-input, \(m\)-output circuit of
length \(L\) into a circuit for the transpose with
\[
L+m-n
\]
additions. Here \(n=9\), \(m=23\), and \(L=14\), yielding a 28-gate
23-to-9 output circuit. A free permutation converts the source's
column-major output-pair convention to ordinary row-major coordinates; in
zero-based notation the permutation is
\([0,3,6,1,4,7,2,5,8]\). Consequently,
\[
13+14+28=55.
\]

\subsection{Why 55 is optimal for this oriented tensor}

\medskip
\noindent\textbf{Proposition 2.}
\textit{For the fixed oriented tensor displayed in this paper, the minimum
sum of the three linear-stage costs in the model of
Section~\ref{sec:model} is 55.}
\medskip

The factor searches give \(C(U)=13\), \(C(V)=14\), and \(C(W)=14\).
For the output side, first prune any dead gates from a hypothetical
\(L\)-gate circuit for \(W^T\). Every one of its 23 inputs is essential and
all nine output forms are nonzero, as is also immediate from the factor
array in Section~\ref{sec:certificate}. The transposition construction
therefore applies to the pruned circuit with all inputs and outputs active,
and gives a circuit for \(W\) of length at most
\(L+9-23=L-14\). Since \(C(W)=14\), necessarily \(L\ge28\). Consequently
\[
C(U)+C(V)+C(W^T)\ge 13+14+28=55,
\]
and the circuits constructed above attain equality.

The lower-bound calculation is deliberately narrow: it certifies the three
linear maps in the displayed orientation, rather than all rank-23 tensors.
Its benefit is that the 55 count is not an artifact of an unfinished
heuristic rescheduling attempt.

\section{Complete straight-line program}
\label{sec:algorithm}

The straight-line program has three stages: form the input linear forms,
multiply corresponding left and right forms, and assemble the output
entries. All formulas are in implementation order. A leading minus sign is
a free sign change.

\subsection{Input networks}

The input stage forms 13 left-side and 14 right-side intermediates:
\begin{center}
\small
\begin{tabularx}{\textwidth}{@{}YY@{}}
\toprule
Left side & Right side\\
\midrule
\(u_1=-A_5+A_2\) & \(v_1=B_8+B_2\)\\
\(u_2=-u_1+A_0\) & \(v_2=B_8+B_5\)\\
\(u_3=u_2+A_1\) & \(v_3=B_2+B_0\)\\
\(u_4=u_3-A_5\) & \(v_4=v_3-B_3\)\\
\(u_5=u_4-A_8\) & \(v_5=B_6+B_0\)\\
\(u_6=u_5-A_1\) & \(v_6=-B_3+B_0\)\\
\(u_7=u_5+A_7\) & \(v_7=v_6-v_2\)\\
\(u_8=u_3-A_4\) & \(v_8=v_4-B_5\)\\
\(u_9=u_8-A_3\) & \(v_9=v_5+B_7\)\\
\(u_{10}=u_9-A_6\) & \(v_{10}=v_9-v_7\)\\
\(u_{11}=u_{10}-A_7\) & \(v_{11}=v_9+B_1\)\\
\(u_{12}=-A_4+A_1\) & \(v_{12}=v_{10}-B_3\)\\
\(u_{13}=-u_{12}+u_{10}\) & \(v_{13}=v_{12}-B_5\)\\
& \(v_{14}=v_{10}+B_4\)\\
\bottomrule
\end{tabularx}
\end{center}

\subsection{Bilinear layer}

Compute the 23 bilinear products:
\begin{center}
\small
\begin{tabularx}{\textwidth}{@{}YY@{}}
\toprule
\multicolumn{2}{c}{Bilinear products}\\
\midrule
\(M_0=u_3v_{10}\) & \(M_{12}=A_5B_7\)\\
\(M_1=u_{10}v_4\) & \(M_{13}=u_9v_3\)\\
\(M_2=A_0v_1\) & \(M_{14}=A_3B_1\)\\
\(M_3=A_7B_4\) & \(M_{15}=A_6B_1\)\\
\(M_4=A_0v_{11}\) & \(M_{16}=A_6B_2\)\\
\(M_5=u_6v_2\) & \(M_{17}=u_1v_5\)\\
\(M_6=u_7B_5\) & \(M_{18}=u_5v_{12}\)\\
\(M_7=A_8B_6\) & \(M_{19}=u_{13}v_6\)\\
\(M_8=u_{11}B_3\) & \(M_{20}=A_1v_{14}\)\\
\(M_9=u_2v_7\) & \(M_{21}=A_4B_4\)\\
\(M_{10}=A_8B_7\) & \(M_{22}=u_8v_8\)\\
\(M_{11}=u_4v_{13}\) & \\
\bottomrule
\end{tabularx}
\end{center}

\subsection{Output network}

The output stage consists of the following 28 additions/subtractions. Read
the left column from top to bottom, followed by the right column from top to
bottom:
\begin{center}
\small
\begin{tabularx}{\textwidth}{@{}YY@{}}
\toprule
\multicolumn{2}{c}{Output straight-line program}\\
\midrule
\(w_1=-M_{18}-M_{10}\) & \(w_{15}=-M_8+w_2\)\\
\(w_2=w_1+M_5\) & \(w_{16}=w_{15}-w_{11}\)\\
\(w_3=-M_7+w_2\) & \(w_{17}=M_4+M_{20}\)\\
\(w_4=M_{11}+w_3\) & \(w_{18}=w_{17}-w_7\)\\
\(w_5=M_0-M_{12}\) & \(w_{19}=M_{14}+M_{21}\)\\
\(w_6=w_5+M_9\) & \(w_{20}=w_{19}+M_{12}\)\\
\(w_7=M_{17}+w_6\) & \(w_{21}=M_3+M_{15}\)\\
\(w_8=M_{16}+w_4\) & \(w_{22}=w_{21}+M_{10}\)\\
\(w_9=M_1+w_8\) & \(w_{23}=M_2-w_4\)\\
\(w_{10}=-M_{13}-M_{11}\) & \(w_{24}=M_{22}+M_{19}\)\\
\(w_{11}=w_{10}+w_9\) & \(w_{25}=w_{24}-M_9\)\\
\(w_{12}=w_3+w_7\) & \(w_{26}=w_{25}-w_9\)\\
\(w_{13}=-M_{19}+w_6\) & \(w_{27}=M_6-M_5\)\\
\(w_{14}=w_{13}+w_{11}\) & \(w_{28}=w_{27}+w_8\)\\
\bottomrule
\end{tabularx}
\end{center}

The nine output entries are free aliases:
\[
\begin{array}{lll}
C_0=w_{12},&C_1=w_{18},&C_2=w_{23},\\
C_3=w_{14},&C_4=w_{20},&C_5=w_{26},\\
C_6=w_{16},&C_7=w_{22},&C_8=w_{28}.
\end{array}
\]
The input stage costs \(13+14=27\) additions, and the output stage costs 28,
for a total of \(27+28=55\). The product stage contains exactly 23
multiplications.

\section{The rank-23 certificate}
\label{sec:certificate}

This section gives two equivalent, fully explicit views of the tensor behind
the circuit. The first favors readability of individual products; the second
is the exact array representation consumed by the verification programs.

\subsection{Expanded products and outputs}
\label{sec:expanded}

The following is the same rank-23 decomposition written without shared input
or output temporaries. It is included for comparison with the compact
algorithm tables in the recent 58-, 59-, and 56-addition papers. This
expanded presentation is not the 55-addition implementation; the
implementation order is Section~\ref{sec:algorithm}.

This section uses one-based matrix entries
\(a_{ij}=A_{3(i-1)+(j-1)}\),
\(b_{ij}=B_{3(i-1)+(j-1)}\), and
\(c_{ij}=C_{3(i-1)+(j-1)}\). The expanded products satisfy
\(p_{r+1}=M_r\) for \(r=0,\ldots,22\).

\begin{lstlisting}[style=certificate]
p01 = (a11 + a12 - a13 + a23) * (b21 + b23 + b31 + b32 + b33)
p02 = (a11 + a12 - a13 - a21 - a22 + a23 - a31) * (b11 + b13 - b21)
p03 = (a11) * (b13 + b33)
p04 = (a32) * (b22)
p05 = (a11) * (b11 + b12 + b31 + b32)
p06 = (a11 - a13 - a33) * (b23 + b33)
p07 = (a11 + a12 - a13 + a32 - a33) * (b23)
p08 = (a33) * (b31)
p09 = (a11 + a12 - a13 - a21 - a22 + a23 - a31 - a32) * (b21)
p10 = (a11 - a13 + a23) * (b11 - b21 - b23 - b33)
p11 = (a33) * (b32)
p12 = (a11 + a12 - a13) * (b31 + b32 + b33)
p13 = (a23) * (b32)
p14 = (a11 + a12 - a13 - a21 - a22 + a23) * (b11 + b13)
p15 = (a21) * (b12)
p16 = (a31) * (b12)
p17 = (a31) * (b13)
p18 = (a13 - a23) * (b11 + b31)
p19 = (a11 + a12 - a13 - a33) * (b23 + b31 + b32 + b33)
p20 = (a11 - a13 - a21 + a23 - a31) * (b11 - b21)
p21 = (a12) * (b21 + b22 + b23 + b31 + b32 + b33)
p22 = (a22) * (b22)
p23 = (a11 + a12 - a13 - a22 + a23) * (b11 + b13 - b21 - b23)

c11 = p01 + p06 - p08 + p10 - p11 - p13 + p18 - p19
c12 = -p01 + p05 - p10 + p13 - p18 + p21
c13 = p03 - p06 + p08 + p11 - p12 + p19
c21 = p01 + p02 + p06 - p08 + p10 - p11 - p13 - p14 + p17 - p19 - p20
c22 = p13 + p15 + p22
c23 = -p02 - p06 + p08 - p10 + p11 - p12 - p17 + p19 + p20 + p23
c31 = -p02 + p08 - p09 + p14 - p17
c32 = p04 + p11 + p16
c33 = p07 - p08 - p11 + p12 + p17 - p19
\end{lstlisting}

\subsection{Complete ternary factor arrays}
\label{sec:factors}

The following gives the tensor factors in compact form. The three blocks are
\(U^T\), \(V^T\), and \(W^T\), separated by a line containing
\texttt{\#}. In each block there are nine rows, one for each row-major
coordinate \(0,\ldots,8\), and each row contains the 23 coefficients for
\(M_0,\ldots,M_{22}\).

Thus the first block gives the coefficients of \(A_0,\ldots,A_8\) in the
left factors, the second gives the coefficients of \(B_0,\ldots,B_8\) in
the right factors, and the third gives the coefficients of
\(C_0,\ldots,C_8\) in the output recombination.

\begin{lstlisting}[style=certificate,basicstyle=\ttfamily\tiny]
1 1 1 0 1 1 1 0 1 1 0 1 0 1 0 0 0 0 1 1 0 0 1
1 1 0 0 0 0 1 0 1 0 0 1 0 1 0 0 0 0 1 0 1 0 1
-1 -1 0 0 0 -1 -1 0 -1 -1 0 -1 0 -1 0 0 0 1 -1 -1 0 0 -1
0 -1 0 0 0 0 0 0 -1 0 0 0 0 -1 1 0 0 0 0 -1 0 0 0
0 -1 0 0 0 0 0 0 -1 0 0 0 0 -1 0 0 0 0 0 0 0 1 -1
1 1 0 0 0 0 0 0 1 1 0 0 1 1 0 0 0 -1 0 1 0 0 1
0 -1 0 0 0 0 0 0 -1 0 0 0 0 0 0 1 1 0 0 -1 0 0 0
0 0 0 1 0 0 1 0 -1 0 0 0 0 0 0 0 0 0 0 0 0 0 0
0 0 0 0 0 -1 -1 1 0 0 1 0 0 0 0 0 0 0 -1 0 0 0 0
#
0 1 0 0 1 0 0 0 0 1 0 0 0 1 0 0 0 1 0 1 0 0 1
0 0 0 0 1 0 0 0 0 0 0 0 0 0 1 1 0 0 0 0 0 0 0
0 1 1 0 0 0 0 0 0 0 0 0 0 1 0 0 1 0 0 0 0 0 1
1 -1 0 0 0 0 0 0 1 -1 0 0 0 0 0 0 0 0 0 -1 1 0 -1
0 0 0 1 0 0 0 0 0 0 0 0 0 0 0 0 0 0 0 0 1 1 0
1 0 0 0 0 1 1 0 0 -1 0 0 0 0 0 0 0 0 1 0 1 0 -1
1 0 0 0 1 0 0 1 0 0 0 1 0 0 0 0 0 1 1 0 1 0 0
1 0 0 0 1 0 0 0 0 0 1 1 1 0 0 0 0 0 1 0 1 0 0
1 0 1 0 0 1 0 0 0 -1 0 1 0 0 0 0 0 0 1 0 1 0 0
#
1 0 0 0 0 1 0 -1 0 1 -1 0 -1 0 0 0 0 1 -1 0 0 0 0
-1 0 0 0 1 0 0 0 0 -1 0 0 1 0 0 0 0 -1 0 0 1 0 0
0 0 1 0 0 -1 0 1 0 0 1 -1 0 0 0 0 0 0 1 0 0 0 0
1 1 0 0 0 1 0 -1 0 1 -1 0 -1 -1 0 0 1 0 -1 -1 0 0 0
0 0 0 0 0 0 0 0 0 0 0 0 1 0 1 0 0 0 0 0 0 1 0
0 -1 0 0 0 -1 0 1 0 -1 1 -1 0 0 0 0 -1 0 1 1 0 0 1
0 -1 0 0 0 0 0 1 -1 0 0 0 0 1 0 0 -1 0 0 0 0 0 0
0 0 0 1 0 0 0 0 0 0 1 0 0 0 0 1 0 0 0 0 0 0 0
0 0 0 0 0 0 1 -1 0 0 -1 1 0 0 0 0 1 0 -1 0 0 0 0
\end{lstlisting}

\section{Exact audit trail}
\label{sec:audit}

The paper's companion repository is \artifactrepo. It contains the frozen
certificate, the exact \texttt{cr58\_cn122} source data, a deterministic
synthesis/replay program, two independently written Python verifiers, and an
independent Node.js \texttt{BigInt} verifier. These artifacts are
self-contained: the Python programs use only the standard library. The
source JSON is retained under the relative path embedded in the frozen
certificate so that regeneration preserves the certificate byte for byte.

From the repository root, run:
\begin{lstlisting}[style=certificate,language=bash]
python3 tools/synthesize_cn122_add55.py \
  --output /tmp/cn122-add55-regenerated.json \
  --summary

cmp /tmp/cn122-add55-regenerated.json \
  search_runs/cn122_add55/certificate.json

python3 tools/verify_cn122_add55_certificate_independent.py \
  search_runs/cn122_add55/certificate.json

python3 search_runs/cn122_add55/verify_independent.py

node search_runs/cn122_add55/verify_independent.js \
  search_runs/cn122_add55/certificate.json
\end{lstlisting}

The synthesis/replay program is also the exhaustive lower-bound prover used
in Section~\ref{sec:reduction}. For each factor map it enumerates every
dependency among basis and target directions, rejects every topological
schedule at the \(d(F)\)-gate floor, and constructs a matching
\(d(F)+1\)-gate circuit. It then expands every gate as an integer coefficient
vector, reconstructs \(U,V,W\), verifies all 729 Brent identities in
\eqref{eq:brent}, and regenerates the frozen certificate byte for byte.
The clean-room Python verifier imports no synthesis code; it independently
decodes the public source tensor, verifies the source and factor hashes,
replays the three task circuits, and checks all 729 identities. The second
independent Python verifier separately audits the source's output-coordinate
permutation and the transposed output map. The Node.js verifier uses
\texttt{BigInt} coefficient arithmetic and a separate implementation.

The exact outputs are:
\begin{lstlisting}[style=certificate]
synthesis/replay:
  factor-map optimality: U=13 (floor 12 impossible),
    V=14 (floor 13 impossible), W=14 (floor 13 impossible)
  U=13, V=14, W=28, total=55
  all 729 Brent identities pass

clean-room Python:
  task gate counts = [13,14,28]
  unit identities = 27
  zero identities = 702
  Brent failures = 0

independent Python:
  U input = 13 PASS
  V input = 14 PASS
  W output task = 28 PASS
  W raw transpose = 28 PASS
  Brent identities checked = 729
  failures = 0

Node.js BigInt:
  circuit maps match certificate = U:true, V:true, W:true
  additions = {U:13,V:14,W:28}
  identities checked = 729
  failures = 0
\end{lstlisting}

The integer Brent check is the proof of correctness. Since the identities
hold over \(\mathbb Z\), all products retain the order
\(L_r(A)R_r(B)\), and the remaining operations are addition and additive
inverse, the same program is valid over every associative ring, including
noncommutative rings. Numerical or randomized tests are not needed for this
conclusion.

\section{Statement on AI-assisted discovery}
\label{sec:ai}

This result arose from a human-directed investigation conducted by a primary AI agent using OpenAI GPT-5.6 Sol, with bounded search, verification, and
review tasks delegated to additional agents. The original objective
was to search for an exact rank-22 decomposition of the \(3\times3\)
matrix-multiplication tensor. During that broader search, the Codex agents
examined recent rank-23 schemes and the mechanisms by which their additive
costs had fallen. The human researcher then redirected the investigation
toward lessons exposed by those experiments: first asking whether a sub-56
circuit could be obtained, and later asking about still lower additive
targets under explicit time and compute budgets.

Within that direction, the primary AI agent selected and implemented the
circuit searches, identified the opportunity in Perminov's
\texttt{cr58\_cn122} tensor, synthesized the 13- and 14-addition factor
circuits, applied the transposition principle to obtain the 28-addition
output circuit, generated the frozen certificate and independent verifiers,
and prepared the initial manuscript. The human researcher supplied the
mathematical objective and acceptance criteria, directed changes of
emphasis, imposed resource limits, and evaluated the significance of
intermediate results.

The tensor decomposition itself was already public and is credited above.
The new claim is the linear schedule and its fixed-tensor optimality
certificate. Because AI systems participated materially in discovery,
programming, verification design, and drafting, every correctness claim is
made reproducible through exact integer data and independently implemented
checks. The disclosure is not offered as a substitute for those checks:
the mathematical certificate is the 729 exact identities, and the operation
count follows from the printed straight-line program.


\begin{thebibliography}{99}

\bibitem{beniamini}
Gal Beniamini, Nathan Cheng, Olga Holtz, Elaye Karstadt, and Oded Schwartz.
\newblock ``Sparsifying the Operators of Fast Matrix Multiplication
Algorithms.''
\newblock arXiv:2008.03759, 2020.

\bibitem{brent}
Richard P. Brent.
\newblock \emph{Algorithms for Matrix Multiplication}.
\newblock Stanford Computer Science Report STAN-CS-70-157, 1970.

\bibitem{burgisser}
Peter Bürgisser, Michael Clausen, and M. Amin Shokrollahi.
\newblock \emph{Algebraic Complexity Theory}.
\newblock Springer, 1997.

\bibitem{karstadt}
Elaye Karstadt and Oded Schwartz.
\newblock ``Matrix Multiplication, a Little Faster.''
\newblock \emph{Journal of the ACM} 67(1):1--31, 2020.

\bibitem{laderman}
Julian D. Laderman.
\newblock ``A Noncommutative Algorithm for Multiplying \(3\times3\)
Matrices Using 23 Multiplications.''
\newblock \emph{Bulletin of the American Mathematical Society}
82(1):126--128, 1976.

\bibitem{martensson-beast}
Erik Mårtensson and Paul Stankovski Wagner.
\newblock ``The Number of the Beast: Reducing Additions in Fast Matrix
Multiplication Algorithms for Dimensions up to 666.''
\newblock In \emph{Proceedings of the Conference on Applied and
Computational Discrete Algorithms}, pages 47--60. SIAM, 2025.
\newblock \url{https://doi.org/10.1137/1.9781611979084.4}.

\bibitem{martensson-rank23}
Erik Mårtensson, Paul Stankovski Wagner, and Joshua Stapleton.
\newblock ``A Rank 23 Algorithm for Multiplying \(3\times3\) Matrices with
an Arithmetic Complexity of 59.''
\newblock arXiv:2601.05272, 2025.

\bibitem{perminov}
Andrew I. Perminov.
\newblock ``A 58-Addition, Rank-23 Scheme for General \(3\times3\) Matrix
Multiplication.''
\newblock arXiv:2512.21980, 2025.

\bibitem{perminov-repo}
Andrew I. Perminov.
\newblock \emph{FastMatrixMultiplication} repository,
\url{https://github.com/dronperminov/FastMatrixMultiplication},
source commit \texttt{98ba522db92b74f1f8c561a78038ff3091356d73}.

\bibitem{smirnov}
Alexey V. Smirnov.
\newblock ``The Bilinear Complexity and Practical Algorithms for Matrix
Multiplication.''
\newblock \emph{Computational Mathematics and Mathematical Physics}
53:1781--1795, 2013.

\bibitem{stapleton}
Joshua Stapleton.
\newblock ``A 60-Addition, Rank-23 Scheme for Exact \(3\times3\) Matrix
Multiplication.''
\newblock arXiv:2508.03857, 2025.

\bibitem{sun}
Yinqi Sun.
\newblock ``An Exact 56-Addition, Rank-23 Scheme for General \(3\times3\)
Matrix Multiplication.''
\newblock arXiv:2604.27645, 2026.

\bibitem{strassen}
Volker Strassen.
\newblock ``Gaussian Elimination Is Not Optimal.''
\newblock \emph{Numerische Mathematik} 13:354--356, 1969.

\end{thebibliography}
\end{document}